\documentclass{ecai}
\usepackage{graphicx}
\usepackage{latexsym}

\usepackage{stmaryrd}
\usepackage{amsfonts}
\usepackage{amssymb}
\usepackage{amsmath}
\usepackage{bbm}
\usepackage{multirow,tabularx}
\usepackage{xspace}
\usepackage{todonotes}
\usepackage{booktabs}
\usepackage{subcaption}
\usepackage{algorithm}
\usepackage{algpseudocode}
\usepackage{marvosym}
\usepackage{hyperref}

\newcommand{\TableHighlight}[1]{\colorbox{lightgray}{#1}}

\newcommand{\fedlearn}{\text{FL}\xspace}

\newcommand{\ESA}{\text{ESA}\xspace}

\newcommand{\nodp}{\text{No-DP}\xspace}

\newcommand{\DPparam}[2]{(#1, #2)-DP\xspace}

\newcommand{\dialfull}{\textbf{EmpatheticDialogue}\xspace}
\newcommand{\newsfull}{\textbf{20NewsGroup}\xspace}

\newcommand{\dialshort}{\textbf{Dial}\xspace}
\newcommand{\newsshort}{\textbf{News}\xspace}

\newcommand{\kmean}{\textbf{K-means}\xspace}
\newcommand{\spectral}{\textbf{Spectral}\xspace}
\newcommand{\greedy}{\textbf{Greedy}\xspace}
\newcommand{\random}{\textbf{Random}\xspace}

\newcommand{\PUR}{\text{Pur.}\xspace}
\newcommand{\RI}{\text{RI}\xspace}
\newcommand{\MI}{\text{MI}\xspace}

\usepackage{ntheorem}

\newtheorem{proposition}{Proposition}

\newtheorem{definition}{Definition} 
\theoremstyle{nonumberplain}

\def\eg{{\em e.g.,}\xspace}
\def\ie{{\em i.e.,}\xspace}

\usepackage{amsmath,amsfonts,bm}

\def\eqref#1{equation~\ref{#1}}

\def\1{\bm{1}}

\def\vb{{\bm{b}}}

\def\vs{{\bm{s}}}

\def\vx{{\bm{x}}}
\def\vy{{\bm{y}}}

\def\mD{{\bm{D}}}

\def\mI{{\bm{I}}}

\def\mM{{\bm{M}}}

\def\mW{{\bm{W}}}

\DeclareMathAlphabet{\mathsfit}{\encodingdefault}{\sfdefault}{m}{sl}
\SetMathAlphabet{\mathsfit}{bold}{\encodingdefault}{\sfdefault}{bx}{n}

\newcommand{\R}{\mathbb{R}}

\begin{document}

\begin{frontmatter}

\title{Fingerprint Attack: Client De-Anonymization in Federated Learning}

\author{\fnms{Qiongkai}~\snm{Xu}\orcid{0000-0003-3312-6825}\thanks{\Letter: qiongkai.xu@unimelb.edu.au}}
\author{\fnms{Trevor}~\snm{Cohn}\orcid{0000-0003-4363-1673}}
\author{\fnms{Olga}~\snm{Ohrimenko}\orcid{0000-0002-9735-0538}} 

\address {The University of Melbourne, Carlton, VIC, Australia}

\begin{abstract}
Federated Learning allows collaborative training without data sharing in settings where participants do not trust the central server and one another. Privacy can be further improved by ensuring that communication between the participants and the server is anonymized through a shuffle; decoupling the participant identity from their data. This paper seeks to examine whether such a defense is adequate to guarantee anonymity, by proposing a novel \emph{fingerprinting attack} over gradients sent by the participants to the server. We show that clustering of gradients can easily break the anonymization in an empirical study of learning federated language models on two language corpora. We then show that training with differential privacy can provide a practical defense against our fingerprint attack.

\end{abstract}

\end{frontmatter}

\begin{abstract}
Federated Learning allows collaborative training without data sharing in settings where participants do not trust the central server and one another. Privacy can be further improved by ensuring that communication between the participants and the server is anonymized through a shuffle; decoupling the participant identity from their data. This paper seeks to examine whether such a defense is adequate to guarantee anonymity, by proposing a novel \emph{fingerprinting attack} over gradients sent by the participants to the server. We show that clustering of gradients can easily break the anonymization in an empirical study of learning federated language models on two language corpora. We then show that training with differential privacy can provide a practical defense against our fingerprint attack.

\end{abstract}

\section{Introduction}

Federated Learning (FL) is a machine learning paradigm that enables collaborative distributed model training without the need to share training data~\cite{mcmahan2016federated}. 
Federated learning usually involves a central server (analyzer) which coordinates training by i) collecting gradient updates from local clients, ii) aggregating the updates; and iii) synchronizing the latest model parameters and sending them back to the clients. 
In such a way, the data from clients never leaves their system, and thus less of their sensitive information is available to the potentially untrusted parties, including the server and other clients.

\begin{figure}[t]
    \centering
    \includegraphics[width=1.0\linewidth]{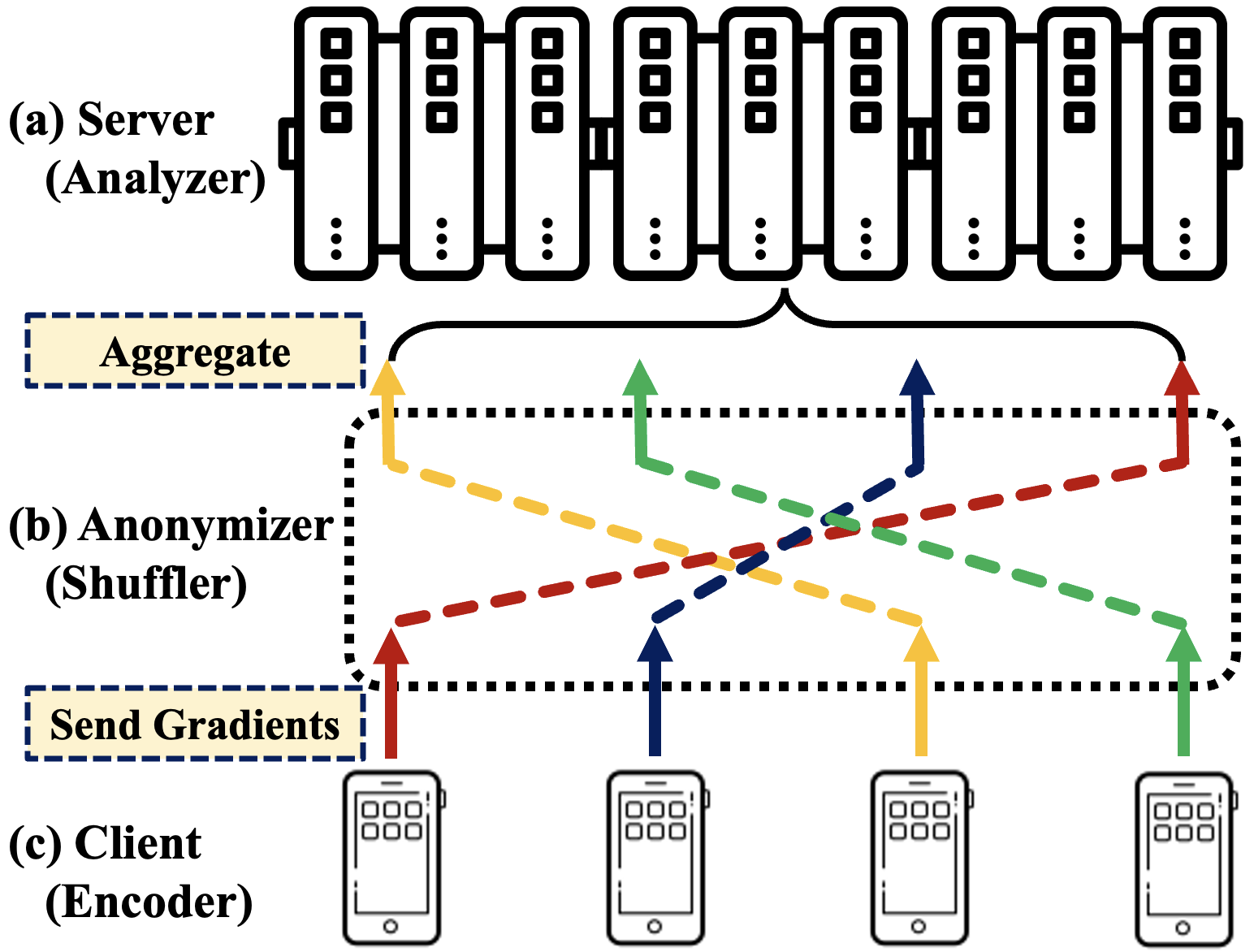}
    \caption{Framework of Federated Learning equipped with Encoder, Shuffler and Analyzer (\ESA)~\cite{bittau2017prochlo}, which are correlated to Server, Anonymizer and Clients. The data collected from clients are gradients of the (language) model indicated by the colored arrows. }
    \label{fig:tease_shuffle_fl}
\end{figure}

Although the training data does not leave clients' local devices, they are still required to communicate to the server key information about the model, namely gradients over their local data from clients to the server. The implicit information in model parameters and their updates have been shown to leak private information through attacks such as membership inference~\cite{melis2019exploiting} and data reconstruction~\cite{gupta2022recovering}. 

Linking data from the same client enables the adversary to perform stronger attacks such as i) combining additional information, \eg user name, home address, and phone numbers that appeared in different batches; and ii) boosting the attack performance by employing multiple gradients from the same source.
Anonymizing client identity is believed to defend against such linkage attacks and amplify the privacy guarantees in distributed and federated learning settings, as data of an individual client is ``concealed'' among the data of other clients~\cite{liu2021flame,erlingsson2019amplification,bittau2017prochlo}. Random data shuffling, performed by a third party other than the server or clients, can be seen as a simple method to anonymize the identities of clients and, hence, enhance privacy in FL, as illustrated in Figure~\ref{fig:tease_shuffle_fl}. A trusted shuffler is placed between clients and the server and operates as follows i) it collects the data packages with model gradients from clients, ii) it removes identities thus anonymizing the providers' identity, and iii) it shuffles data and sends them to the central server. 
By delinking data from the same client the Shuffle module provides a defense against attacks that use multiple gradients over time. 
Intuitively, shuffling in FL limits the effectiveness of many attacks, as the server can only exploit single gradients rather than a full sequence of client updates from a training run. 

Our work attempts to challenge the anonymization guarantees in the Shuffle-FL algorithm through a novel fingerprinting attack. In fingerprinting attack, a curious-but-honest\footnote{The server honestly follows the FL protocol but it is curious to learn the composition of the clients' datasets.} service provider records the gradients from all clients in the training process. We posit that the gradients from a client contain substantial information that is unique to that client, and thus provides a unique \emph{fingerprint}.
We propose an attack based on clustering and greedy match algorithms over pairs of gradients, in order to recover which data updates came from the same clients. 
We evaluate the effectiveness of our fingerprinting attack through extensive experiments on FL language modeling, showing substantially above-chance performance, and in some settings, perfect linking. As a defense, we apply differential privacy on the gradients before they are collected by the shuffler or analyzer. Our study shows that deferentially private gradients reduce the performance of fingerprinting attacks, although at a cost to model utility and training efficiency. 

We summarize our contributions as follows:

\begin{itemize}
    \item To the best of our knowledge, we are the first to propose a fingerprinting attack against the shuffler in the federated learning setting. The shuffled gradients could be grouped by greedy matching and clustering algorithms and thus traced to the same clients. 
    \item We empirically demonstrate the feasibility of fingerprinting attacks on federated training when training a language model. 
    \item We explore differential privacy as a defense and empirically show its effectiveness in defending against fingerprint attack, while providing a privacy-utility trade-off.
\end{itemize}

\section{Related Work}

\paragraph{Federated Learning}
Federated learning is a framework for collaboratively training machine learning models~\cite{kairouz2021advances,mcmahan2016federated,konecnyfederated}. The general federated learning framework is composed of i) clients, who train local models using their data and periodically communicate the parameter updates to the server, and ii) a server, which aggregates received model updates and synchronizes the new parameters among the clients across several rounds of training. 
Federated learning has many applications~\cite{mcmahan2016federated} including those where (1) training of advanced deep neural network (DNN) requires a high volume of data~\cite{zhang2018survey,goodfellow2016deep} that is unlikely to be owned by a single party; (2) the data cannot leave client's devices, for example, when training a diagnostic model across multiple institutions to predict clinical outcomes in patients with COVID-19 while maintaining data anonymity~\cite{dayan2021federated}. 
Language modeling is one of the fundamental tasks in Natural Language Processing and FL for language modeling recently attracted attention in academia and industry~\cite{passban-etal-2022-training,weller-etal-2022-pretrained,chen2019federated}. 

\paragraph{Attacks and Defenses in FL}
Several security and privacy challenges have been identified in adapting federated learning.

The first concern is the impact of malicious participants on the model learning who can backdoor the model to have a specific prediction when a trigger is given in the input~\cite{bagdasaryan2020backdoor}. A series of strategies are proposed to eliminate the confounding contributions from malicious clients, such as certifiably robust models against backdoor in FL~\cite{xie2021crfl} and Krum~\cite{blanchard2017machine} against Byzantine generals problem~\cite{lamport1982byzantine}.

A second critical concern is \textit{membership inference}~\cite{shokri2017membership} that the attackers can determine if data was utilized in the federated model training or not~\cite{melis2019exploiting}.
A second critical concern is whether the client's local data will be disclosed to other parties in training. 
More ambitious are \textit{data reconstruction} attacks, which aim to recover samples used in training. Methods for inverting gradients were proposed to reconstruct the exact training image from the first linear layer of deep neural models~\cite{geiping2020inverting}. 
The following work recovered the private texts by first identifying the set of used words and then directly reconstructing sentences based on beam search~\cite{gupta2022recovering}. These attacks utilize the model parameters and their recent updates to infer the training data.

Common methods of defense are to perturb the parameters or model updates using differential privacy~\cite{abadi2016deep,lyu2020differentially}. Shuffle models were proposed to enhance the privacy protection in FL~\cite{girgis2021shuffled,liu2021flame}, as individual data items are shuffled and thus anonymously hidden in a larger batch of data which increases the difficulty of discriminating their usage~\cite{bittau2017prochlo}.

Our work belongs to the second challenge and the proposed fingerprinting attack serves as a new threat to FL, specifically aiming at disabling the privacy amplification of the shuffle module.

\section{Fingerprinting Attack}
In this section, we first formulate the federated learning framework with a shuffle module. Then, we describe our proposed fingerprinting attack.
\subsection{Preliminaries}
\paragraph{Federated Learning.}
Federated Learning (\fedlearn) trains a machine learning model $f(\vx; \Theta)$ using data of multiple clients or silos. For each iteration $t\in \llbracket 1..T \rrbracket$, the participating client $k\in \llbracket 1..K \rrbracket$ calculates the gradients of the model based on subset samples of their own data $\{\vx^{k}_t\}$,
\begin{equation}
    \theta^{k}_t \leftarrow \nabla \mathcal{L}(\vx^{k}_t; \Theta_t)
\end{equation}
which often involves performing several iterations of mini-batch SGD locally on the client.
Then, the server aggregates the gradients by averaging the updates from the clients,
\begin{equation}
    \Theta_{t+1} \leftarrow \Theta_t - \lambda \cdot \text{Avg} (\{\theta^{k}_t\}). 
    \label{eqn:fl_avg}
\end{equation}
The procedures for averaging parameters can vary, and we primarily use FedAvg~\cite{mcmahan2017communication}.
The updated parameters are then distributed to the clients, and the process is repeated for several epochs, until convergence. 

\paragraph{Shuffle Module.} Encode, Shuffle and Analyze (\ESA)~\cite{bittau2017prochlo} is a framework proposed for amplifying privacy protection by adding a shuffle module as an anonymizer between client-server communication. 
Shuffle module was also proved to provide a better privacy guarantee in the federated learning setting when combined with differential privacy~\cite{liu2021flame}. 
The shuffle model $\mathbb S$ anonymizes the client identities by permuting the data for analysis, \ie gradients sent from clients to the server in our case:
\begin{equation}
    \mathbb S( \langle \theta^{k} \rangle )= \langle \theta^{p(k)} \rangle,
\end{equation}
where $\langle \cdot \rangle$ is an ordered sequence and $p(\cdot)$ indicates a permutation function on $\llbracket 1..K \rrbracket$. 
The shuffle breaks the link between individual clients and their data. Moreover, the data is mixed with data from other clients.
Accordingly, the server should no longer be able to exploit the information that several gradients come from the same source, and their link to the individual client. 

The shuffle has no effect on the `honest' computation of the server, \ie the server can still perform the aggregated update in Equation~\ref{eqn:fl_avg} based on the shuffled results,
\begin{equation}
    \Theta_{t+1} \leftarrow \Theta_t - \lambda \cdot \text{Avg} (\{\theta^{p(k)}_t\}) 
\end{equation}
to produce an identical result as permutation does not affect the average value of a set, 
\begin{equation}
    \text{Avg} (\{\theta^{p(k)}_t\})=\text{Avg} (\{\theta^{k}_t\}).
\end{equation}

\subsection{Fingerprinting Attack against Anonymization}

We hypothesize that the linear layer gradients from the same clients should be ``similar'' to each other as i) each client possesses data in specific domains which decides the distribution of hidden representation, and ii) these vectors are the main factors in gradient calculation of corresponding linear layers. If that is the case, the attacker can use gradients to group data that comes from the same client. In this section, we first provide an analysis on the intuition of fingerprinting attacks. Then, we describe our fingerprinting attack methods, based on standard clustering techniques and a simple greedy match algorithm. 
We note that though clustering methods have been used to enhance the training of FL\cite{criado2022non,sattler2020clustered}, we use clustering as an adversarial tool to perform fingerprinting attacks.

\subsubsection{Attack Intuition}

Inspired by a data reconstruction attack~\cite{geiping2020inverting}, we note that the gradients with respect to the parameters of a linear layer in a neural network can be used to recover the inputs to that layer.   

\begin{definition}
Given a linear layer,
\begin{equation}
    \vy = \mW \vx + \vb,
    \label{eq:linear_layer}
\end{equation}
where $\mW\in \R^{M\times N}$, $\vx = (x_1, \cdots, x_N)^\top$ and $\vy = (y_1, \cdots, y_M)^\top$ are the weight matrix, the inputs and the outputs respectively. 
The gradient of $\mW$ with regard to loss $\mathcal L$ is defined as 
\begin{equation}
    \Delta\mW \triangleq \frac{\partial \mathcal L}{\partial \mW}.
\end{equation}
\end{definition}
\begin{proposition} \label{thm:gradient_base}
The gradient $\Delta\mW$ of a linear layer is associated with its input $\vx$,
\begin{equation} \label{eq:deltaw}
    \frac{\partial \mathcal L}{\partial \mW} =  \frac{\partial \mathcal L}{\partial \vb}  \cdot \vx^\top. 
\end{equation}
\end{proposition}
The proposition is derived by the chain rule of derivation and the fact that
\begin{equation}
    \frac{\partial \mathcal L}{\partial \vy}= \frac{\partial \mathcal L}{\partial \vb}.
\end{equation}
Due to the connection between $\Delta\mW$ and $\vx$, we hypothesize that the gradients $\Delta\mW$ from the same clients are similar to each other, given that their training data possesses similar textual patterns, \eg topics or writing styles. We will show how to measure the similarity of the gradients in the following discussion.

\subsubsection{Distance Measurement}

We rely on a distance metric $\mathbb D(\theta, \theta')$ to capture the relation between the model gradients $\theta$ and~$\theta'$. Note that gradients from linear layers $\Delta \mW$ are a subset of overall gradients $\theta$. 
Inspired by Gradient Inversion~\cite{geiping2020inverting}, we consider the linear layers to conduct fingerprinting attacks. As the Transformer~\cite{vaswani2017attention} is the current dominant model architecture in NLP, we focus our experiments on this architecture, for FL fine-tuning of the GPT-2~\cite{radford2019language} language model. 
All parameters are concatenated into a vector and then normalized. 
Euclidean distance is considered in clustering algorithms and negative cosine similarity is utilized in Greedy Match. 
The rationality of negative cosine similarity is that it is proportional to the Euclidean distance between the normalized vectors.

\subsubsection{Naive Clustering} The inputs to the clustering are the complete collection of  $K \times T$ gradient vectors as computed by $K$ clients performing FL for $T$ epochs.
We use gradients of linear layers $\Delta \mW^{k}$ as a subset of $\theta^{k}$ to construct the features. Clustering aims to assign the close vectors to the same group. 
To verify our design, we consider two representative clustering methods, K-means Clustering (\kmean)~\cite{lloyd1982least} and Spectral Clustering (\spectral)~\cite{ng2001spectral}.
\kmean finds cluster centers that minimize the intra-class variance, which is iteratively optimized by calculating cluster centroids and data assignments to clusters. 
\spectral performs dimensionality reduction on the similarity matrix of the data before clustering.

\subsubsection{Step-wise Greedy Match} 
Intuitively, gradients from the nearest steps possess the most signal, as the model has fewer parameter updates between neighboring steps. We propose to trace the alignments of $t\in \llbracket 1 .. T\rrbracket$ training steps with $t+1$ steps to group the data from clients.
For two neighboring epochs in the training sequence, we find the optimal alignment between the $K$ clients' data from step $t$ and step $t+1$. 
The similarity between the client gradients from two steps $\theta^{(t)}$ and $\theta^{(t+1)}$ is measured by the distance matrix $\mD \in \R^{K\times K} $,
$\mD_{ij} = \mathbb{D}(\theta^{(t)}_i, \theta^{(t+1)}_j)$ where $i,j \in \llbracket 1 .. K\rrbracket$ are party identifiers. 
Greedy selection attempts to find the best pairings, by minimizing the distance for each adjacent time step.
This formulation can be solved over the time series of parameter vectors to find a step-wise globally optimal alignment through solving a Linear Sum Assignment Problem using the Hungarian algorithm~\cite{kuhn1955hungarian}, where $\mM$ is a matrix with each value $\mM_{i,j}$ indicating a 0-1 assignment between gradients $i$ and $j$.
\begin{align}
    & \min_{\mM}  \sum_i \sum_j \mD_{i, j} \mM_{i, j} \\
    \nonumber \text{s.t. } & \forall i, \sum_j \mM_{i, j}=1; \forall j, \sum_i \mM_{i, j}=1; \\
    & \forall i, j, \mM_{i, j} \in \{0, 1\}. \nonumber
\end{align}
The distance is defined as the negative cosine similarity of two vectors,
\begin{equation}
    \mathbf{D}_{i, j}=1-\text{Sim}(\theta^{(t)}_{i}, \theta^{(t+1)}_{j}),
\end{equation}
which is proportional to the Euclidean distance between two normalized gradients.

\section{Attack Experiments}
We conduct experiments on language modeling to show the effect of fingerprinting attacks.\footnote{The code and its guideline are available at \url{https://github.com/xuqiongkai/FingerprintAttack_on_FL}.}

\subsection{Experimental Settings}

\paragraph{Datasets.}

We evaluate the fingerprinting attack on the language modeling task using two datasets: \newsfull (\newsshort)~\cite{lang1995newsweeder} and \dialfull (\dialshort)~\cite{rashkin-etal-2019-towards}. 
These datasets are most commonly used for text classification, rather than language modeling, but we use them here as they contain natural data divisions that are a good match for typical FL scenarios.
For \newsshort, we distributed the data to 20 clients, where each client accesses text samples in a single topic. 
For \dialshort, we include the 70 speakers with more than 288 utterances from the original dataset, with each speaker comprising a client. 
The statistics of the datasets are shown in Table~\ref{tab:dataset}.
\begin{table}
    \centering
    \caption{Statistic of \newsfull (\newsshort) and \dialfull (\dialshort), with number of samples in train and valid sets. The total number of used samples equals to (\#Train+\#Valid)$\times$\#Clients.}
    \begin{tabular}{l cccc}
      \toprule
    
Dataset &  \#Train & \#Valid & \#Clients & Total\\
    \midrule
\newsshort &  512 & 64 & 20 & 11,520 \\
\dialshort &  256 & 32 & 70 & 20,160 \\
    \bottomrule
 \end{tabular}

    \label{tab:dataset}
\end{table}

\paragraph{Federated Learning.} We simulate Federated Learning on an Nvidia A100 server based on FLSim\footnote{\url{https://github.com/facebookresearch/FLSim}}. All language models are initialized by loading a pre-trained GPT-2 model, the learning rate of the server is selected based on  preliminary experiments, and the learning rate of clients' local training is set to 0.1 in all our experiments. We set the maximum sentence length to 40 tokens due to the limitation of our computational resources.
Please see Appendices~\ref{appendix:pre_train} and~\ref{appendix:fl_setting} for further details. We use stochastic gradient descent (SGD) optimizer without momentum to update the model parameters for each client.\footnote{Momentum or other gradient smoothing methods would necessitate special treatment in the attack, and would otherwise cause false positive matches.}

\paragraph{Language Model.} As the Transformer~\cite{vaswani2017attention} is the current dominant model architecture in NLP, we focus our experiments on this architecture, for FL fine-tuning of the GPT-2~\cite{radford2019language} language model. We customized a smaller GPT model with 4 Transformer layers and pre-trained it on WikiText101. Please see more details about the model in Appendix~\ref{appendix:pre_train}. 
Our experiments, focus on the linear layers in feedforward modules, which are denoted as fully connected layers (FC) and projection layers (Proj). 
All parameters are concatenated into a vector and then normalized.

\paragraph{Evaluation.} 
We evaluate the attack performance using standard evaluation metrics for clustering~\cite{manning2008introduction}. 
\begin{itemize}
\item \textbf{Purity Score} (\PUR)~\cite{manning2008introduction} measures the proportion of the dominant class over all clusters.
\item \textbf{Rand Index} (\RI)~\cite{rand1971objective} measures the percentage of the correct decision pairs between all data points.
\item \textbf{Mutual Information} (\MI)~\cite{pfitzner2009characterization} is a measurement of the information shared between a clustering result and the ground truth. 
\end{itemize}

\begin{table*}[t]
\caption{The comparison of fingerprint attack on federated learning with various number of clients (\# Clients) on \newsfull based on purity, rand-index and mutual information (higher means better attack success). } 
\begin{center}
\begin{tabular}{ c c ccc }
\toprule
 & \random & \kmean & \spectral & \greedy \\
\cmidrule(r){2-5}
\# Clients & \PUR / \RI / \MI & \PUR / \RI / \MI & \PUR / \RI / \MI & \PUR / \RI / \MI \\
\midrule
3  & 0.458 / 0.563 / 0.074 & 0.667 / 0.699 / 0.547 & \TableHighlight{1.000 / 1.000 / 1.099} & \TableHighlight{1.000 / 1.000 / 1.099} \\\midrule
5  & 0.351 / 0.689 / 0.192 & 0.400 / 0.481 / 0.500 & 0.960 / 0.971 / 1.501 & \TableHighlight{1.000 / 1.000 / 1.609} \\\midrule
10 & 0.260 / 0.827 / 0.484 & 0.200 / 0.264 / 0.325 & 0.850 / 0.960 / 2.031 & \TableHighlight{1.000 / 1.000 / 2.303} \\\midrule
20 & 0.206 / 0.909 / 0.933 & 0.100 / 0.138 / 0.199 & 0.245 / 0.785 / 0.766 & \TableHighlight{1.000 / 1.000 / 2.996} \\

\bottomrule
\end{tabular}
\end{center}
\label{tab:client_scale_news}
\end{table*}

\begin{table*}[t]
\caption{The comparison of fingerprint attack on federated learning with various number of clients (\# Clients) on \dialfull based on \PUR/\RI/\MI. The number of clients is between 5 to 70.}
\begin{center}
\begin{tabular}{ c c ccc }
\toprule
 & \random & \kmean & \spectral & \greedy \\
\cmidrule(r){2-5}
\# Clients & \PUR / \RI / \MI & \PUR / \RI / \MI & \PUR / \RI / \MI & \PUR / \RI / \MI \\
\midrule
3  & 0.458 / 0.563 / 0.074 & 0.400 / 0.359 / 0.074 & \TableHighlight{0.633 / 0.607 / 0.342} & 0.500 / 0.600 / 0.181 \\\midrule
5  & 0.351 / 0.689 / 0.192 & 0.280 / 0.282 / 0.132 & 0.400 / 0.676 / 0.341 & \TableHighlight{0.580 / 0.770 / 0.584} \\\midrule
10 & 0.260 / 0.827 / 0.484 & 0.300 / 0.528 / 0.503 & 0.500 / 0.853 / 1.121 & \TableHighlight{0.510 / 0.881 / 1.123} \\\midrule
20 & 0.206 / 0.909 / 0.933 & 0.305 / 0.776 / 0.976 & 0.450 / 0.911 / 1.592 & \TableHighlight{0.500 / 0.939 / 1.761} \\\midrule
40 & 0.170 / 0.954 / 1.486 & 0.280 / 0.875 / 1.382 & 0.425 / 0.959 / 2.143 & \TableHighlight{0.445 / 0.967 / 2.305} \\\midrule
70 & 0.148 / 0.973 / 1.984 & 0.250 / 0.903 / 1.728 & 0.449 / 0.976 / 2.745 & \TableHighlight{0.497 / 0.983 / 2.983} \\

\bottomrule
\end{tabular}
\end{center}

\label{tab:client_scale_dial}
\end{table*}

\subsection{Results}
\paragraph{Comparison of fingerprint attack methods.}
We run 10 epochs of federated learning of language model on \newsfull and \dialfull, involving all 20 and 70 clients respectively. We report the performance of fingerprinting attacks in Tables~\ref{tab:client_scale_news} and~\ref{tab:client_scale_dial}. The proposed fingerprinting attacks achieve better clustering results than random baselines. \spectral consistently works better than \kmean on both datasets, confirming the utility of its additional dimensionality reduction step.
\greedy works the best among all our methods, achieving perfect grouping results on \newsfull. 
We attribute this to its explicit formulation enforcing a balanced partitioning of the data. 
With more clients in FL, the fingerprinting attack becomes more difficult as the attacker needs to consider more combinations of potential client-gradient mappings. This is reflected by the decrease in clustering performance with the growth of client numbers. Nonetheless, \spectral and \greedy still maintain high attack performance on both datasets.

\begin{figure*}
     \centering
     
     \begin{subfigure}[b]{0.35\textwidth}
         \centering
         \includegraphics[width=\textwidth]{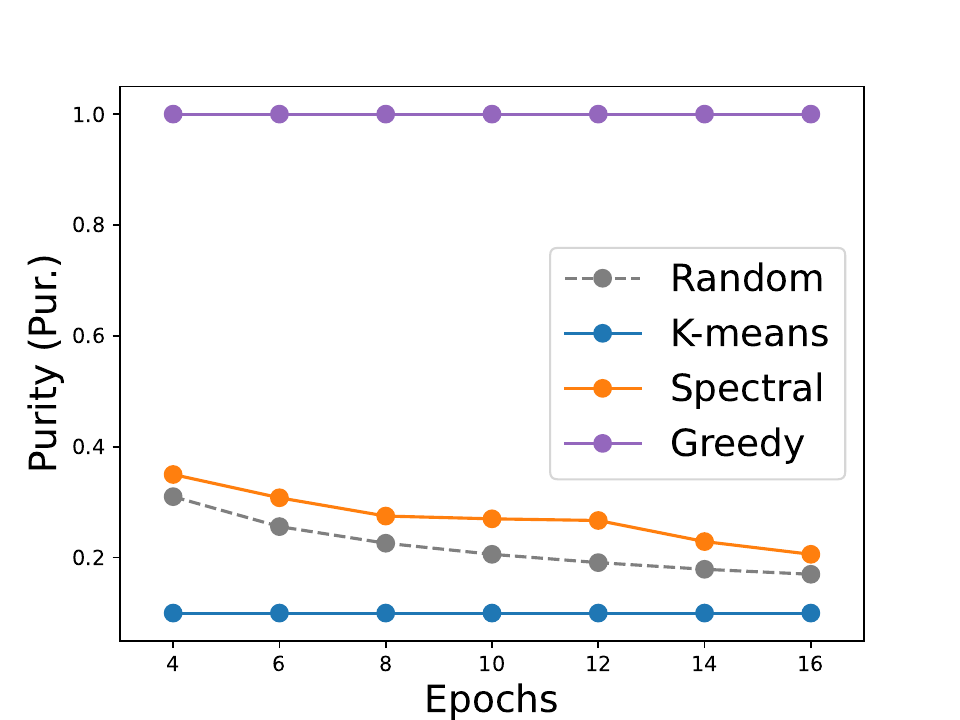}
         \caption{\newsshort (\PUR)}
         \label{fig:epoch_news_PUR}
     \end{subfigure}
     \hspace{-6mm}
     \begin{subfigure}[b]{0.35\textwidth}
         \centering
         \includegraphics[width=\textwidth]{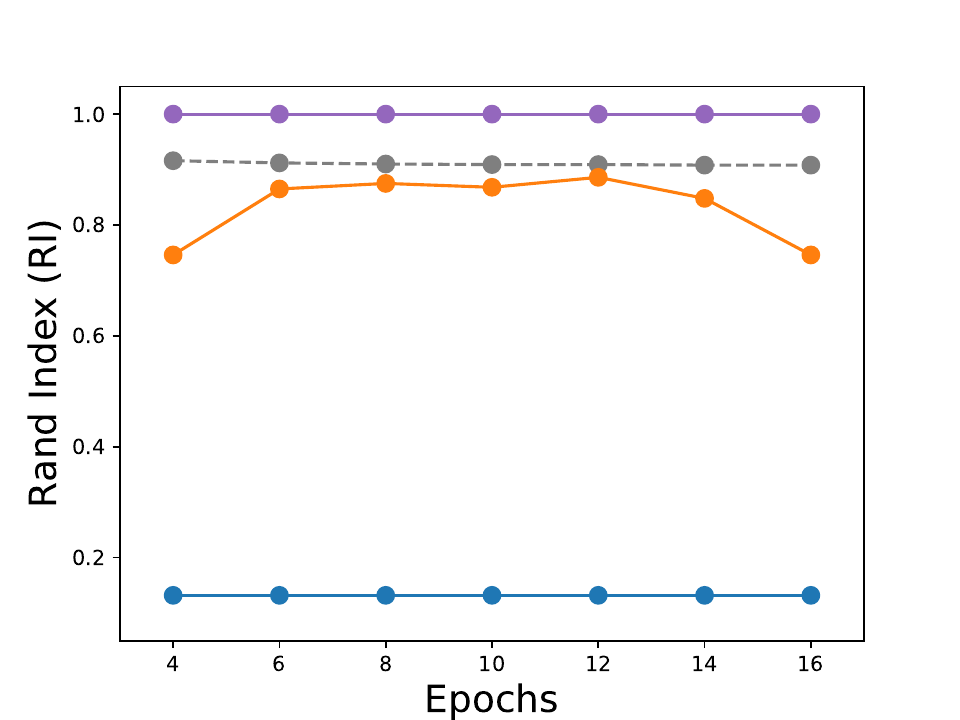}
         \caption{\newsshort (\RI)}
         \label{fig:epoch_news_RI}
     \end{subfigure}
     \hspace{-6mm}
     \begin{subfigure}[b]{0.35\textwidth}
         \centering
         \includegraphics[width=\textwidth]{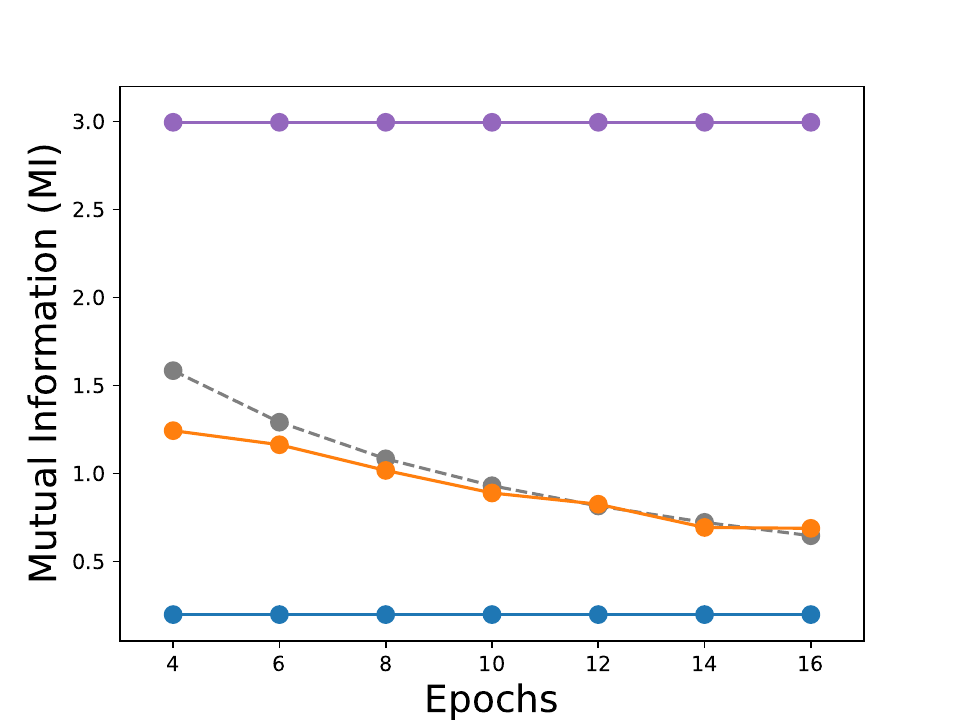}
         \caption{\newsshort (\MI)}
         \label{fig:epoch_news_MI}
     \end{subfigure}
     
     \begin{subfigure}[b]{0.35\textwidth}
         \centering
         \includegraphics[width=\textwidth]{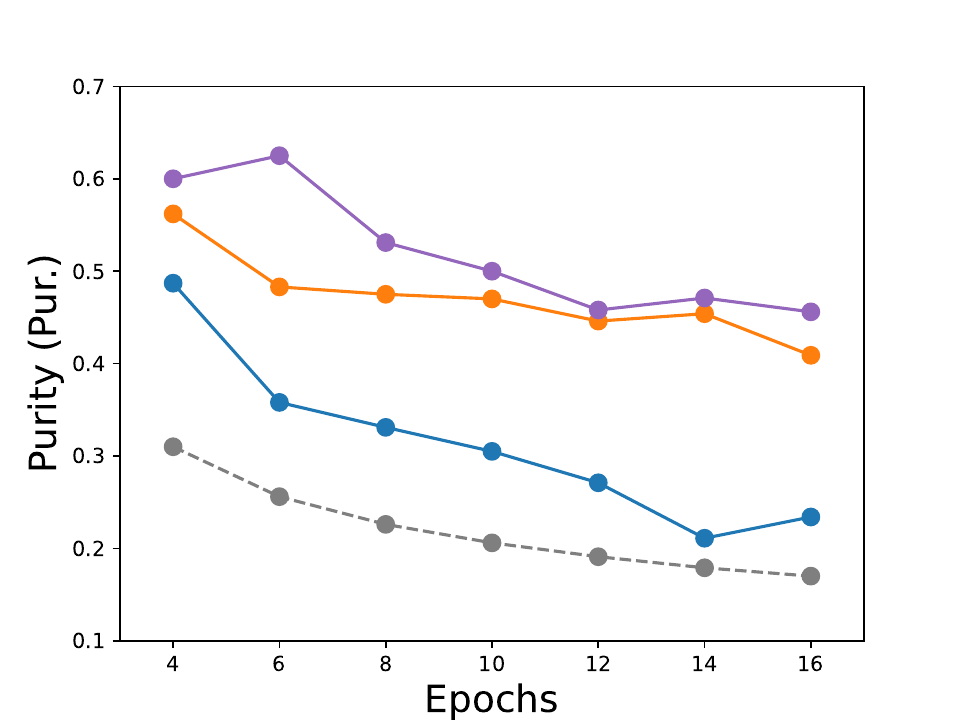}
         \caption{\dialshort (\PUR)}
         \label{fig:epoch_dial_PUR}
     \end{subfigure}
     \hspace{-6mm}
     \begin{subfigure}[b]{0.35\textwidth}
         \centering
         \includegraphics[width=\textwidth]{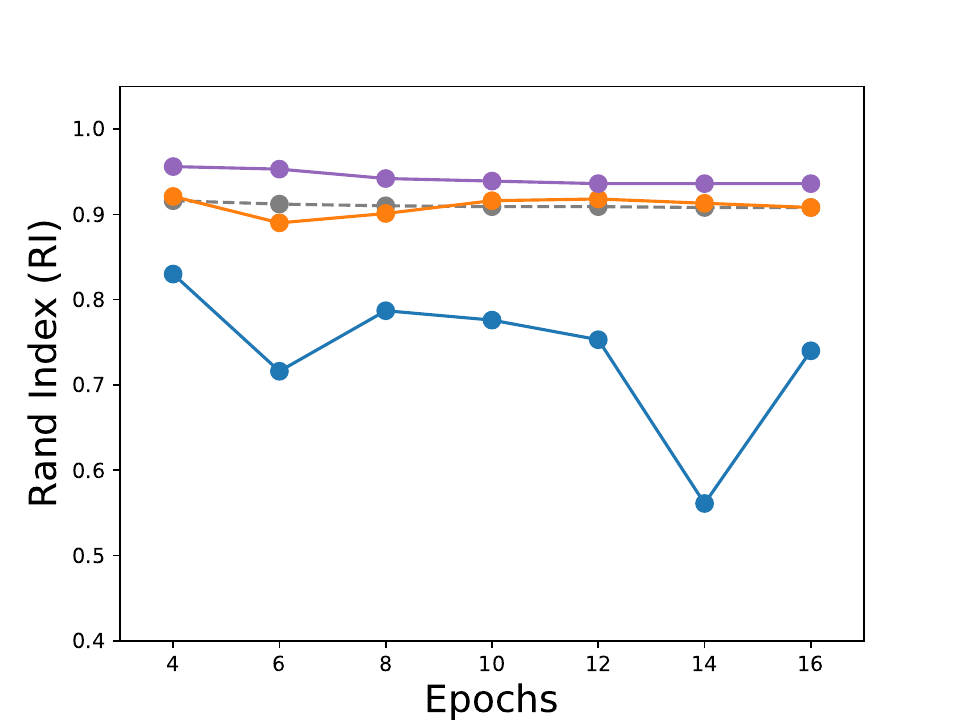}
         \caption{\dialshort (\RI)}
         \label{fig:epoch_dial_RI}
     \end{subfigure}
    \hspace{-6mm}
     \begin{subfigure}[b]{0.35\textwidth}
         \centering
         \includegraphics[width=\textwidth]{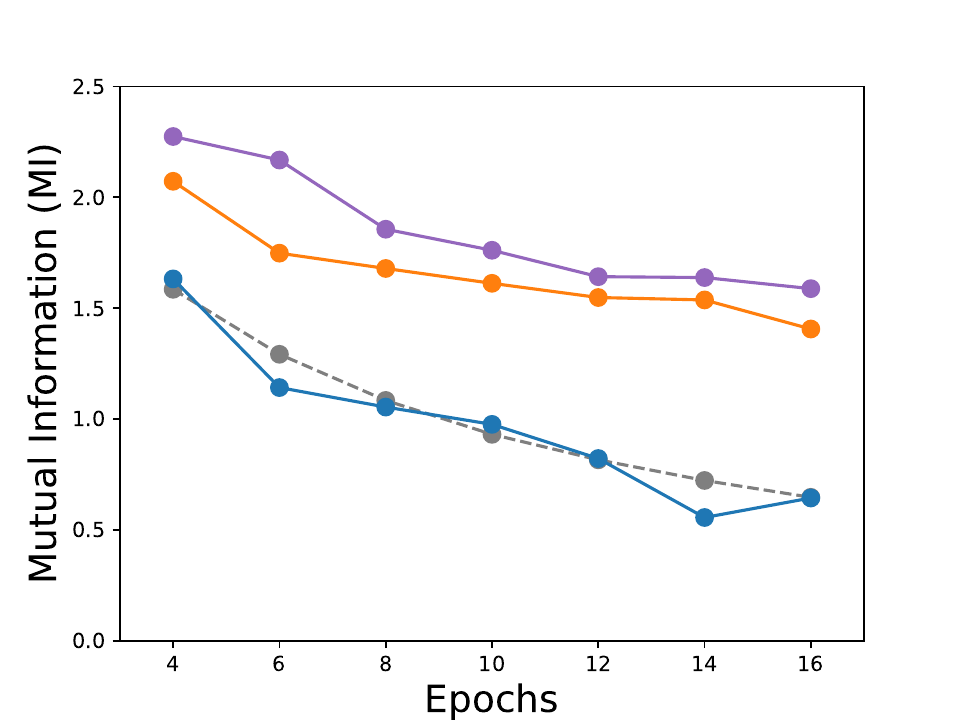}
         \caption{\dialshort (\MI)}
         \label{fig:epoch_dial_MI}
     \end{subfigure}
     
        \caption{The comparison of fingerprinting methods (\kmean, \spectral, and \greedy) with 4 to 16 epochs on \newsshort and \dialshort using Purity (\PUR). Rand Index (\RI) and Mutual Information (\MI).
        }
        \label{fig:comp_epoch}
\end{figure*}

\paragraph{Impact of the number of training epochs.}
During training, the parameters of the language model are continuously updated and synchronized for each epoch. We investigate the influence of the model dynamic by varying the training epochs in our FL experiments. The fingerprinting attack performance on \newsshort and \dialshort is illustrated in Figure~\ref{fig:comp_epoch}. There is a clear trend of the attack becoming more difficult with the increasing number of epochs. We note that larger epoch gaps mean more difference in model states by the same client; this lead to more divergent intermediate representations $\vx$ given as inputs to the attacked linear layer.  
The \newsfull dataset shows a clear difference in efficacy of the methods, with the random performance from \kmean and perfect performance for \greedy, uniformly across all experiment sizes.

\paragraph{Comparison of feature construction.} We are interested in the impact of feature construction on the success of the fingerprinting attack. We compare $\Delta \mW$ from the feedforward modules in Transformer, using the fully connected layers (FC) and the projection layers (Proj). All the variations of feature combination and selection of layers are effective in our attack. Combining both features achieves the best performance, as demonstrated in Table~\ref{tab:main_feature_combine}. The results for layer selection are reported in Table~\ref{tab:compare_layer}. We have not found a layer that outperforms others by a significant margin; accordingly we use only the first layer in our primary experiments. 

\begin{table}
\caption{The comparison of features for fingerprinting attacks on \newsfull and \dialfull.}
\begin{center}
\begin{tabular}{ c cc }
\toprule
 & \spectral & \greedy \\
\cmidrule(r){2-3}
Feature & \PUR / \RI / \MI & \PUR / \RI / \MI \\
\midrule

FC & 0.270 / 0.905 / 1.102 & 1.000 / 1.000 / 2.996 \\
Proj & 0.245 / 0.795 / 0.699 & 1.000 / 1.000 / 2.996 \\\midrule
Both & 0.275 / 0.900 / 1.034 & 1.000 / 1.000 / 2.996 \\
\bottomrule
\\[-0.3em]
\multicolumn{3}{c}{(a) \newsfull} \\
\\[-0.3em]
\toprule
 & \spectral & \greedy \\
\cmidrule(r){2-3}
Feature & \PUR / \RI / \MI & \PUR / \RI / \MI \\
\midrule
FC & 0.390 / 0.916 / 1.469 & 0.465 / 0.935 / 1.690 \\
Proj & 0.405 / 0.917 / 1.504 & 0.445 / 0.935 / 1.698 \\\midrule
Both & 0.465 / 0.912 / 1.610 & 0.500 / 0.939 / 1.761 \\
\bottomrule
\\[-0.3em]
\multicolumn{3}{c}{(b) \dialfull} \\
\end{tabular}
\end{center}

\label{tab:main_feature_combine}
\end{table}

\begin{table}
\caption{The comparison of fingerprinting attacks on different layers $\{1, 2, 3, 4\}$ in transformer models. Both gradients of the connected layer (FC) and projection layer (Proj) are used. Purity (\PUR), Rand Index (\RI) and Mutual Information (\MI) are reported.}
\begin{center}
\begin{tabular}{ c cc }

\toprule
  & \spectral & \greedy \\
\midrule
Layer & \PUR / \RI / \MI & \PUR / \RI / \MI\\
\midrule
1 & 0.275 / 0.900 / 1.034 & 1.000 / 1.000 / 2.996 \\
2 & 0.280 / 0.883 / 0.908 & 1.000 / 1.000 / 2.996 \\
3 & 0.255 / 0.874 / 0.881 & 1.000 / 1.000 / 2.996 \\
4 & 0.285 / 0.881 / 0.962 & 1.000 / 1.000 / 2.996 \\
\bottomrule
\\[-0.3em]
\multicolumn{3}{c}{(a) \newsfull} \\
\\[-0.2em]
\toprule
   & \spectral & \greedy \\
\midrule
Layer & \PUR / \RI / \MI & \PUR / \RI / \MI\\
\midrule
1 & 0.465 / 0.912 / 1.610 & 0.500 / 0.939 / 1.761 \\
2 & 0.465 / 0.922 / 1.588 & 0.535 / 0.941 / 1.813 \\
3 & 0.465 / 0.927 / 1.628 & 0.520 / 0.942 / 1.869 \\
4 & 0.445 / 0.928 / 1.583 & 0.490 / 0.939 / 1.777 \\
\bottomrule
\\[-0.3em]
\multicolumn{3}{c}{(b) \dialfull} \\
\end{tabular}
\end{center}
\label{tab:compare_layer}
\end{table}

\section{Defense on Fingerprinting Attack}

\begin{figure*}[t]
  \captionof{table}{Clustering performance (\PUR/\RI/\MI) on gradients with DP-SGD on \newsfull, using different clipping bounds $\mathcal{C}$ and noise multipliers $\sigma$. Target delta $\delta=10^{-4}$ and corresponding $\varepsilon$ are used to demonstrate the DP budget. We highlight the results according to the attack performance. The baseline performance of FL without DP-SGD (No-DP) is reported in the captions of sub-tables.}
  \begin{minipage}[b]{0.99\textwidth}
    \centering
    \scalebox{1.}{
    \begin{tabular}{c  c c c}
    \toprule
         \multicolumn{1}{r}{}   & $\sigma=0.5$ & $\sigma=1.0$ & $\sigma=1.5$ \\ 
        $\mathcal{C}$ & \DPparam{10.87}{$10^{-4}$} & \DPparam{2.216}{$10^{-4}$} & \DPparam{0.935}{$10^{-4}$}\\ 
    \midrule
        50 & \colorbox{gray!70}{0.685/0.959/2.283} & \colorbox{gray!40}{0.220/0.915/1.033} & \colorbox{gray!40}{0.205/0.913/0.930} \\
        100 & \colorbox{gray!40}{0.215/0.914/0.990} & \colorbox{gray!40}{0.225/0.914/1.007} & \colorbox{gray!40}{0.210/0.914/0.988} \\
        200 & \colorbox{gray!40}{0.210/0.914/0.971} & \colorbox{gray!40}{0.245/0.916/1.087} & \colorbox{gray!40}{0.245/0.915/1.054} \\
    \bottomrule
        \\[-0.3em]
        \multicolumn{2}{r}{(a) \greedy} & \multicolumn{2}{l}{w. \nodp: 1.000 / 1.000 / 2.996} \\
        \\[-0.3em]
        \hline
        
    \toprule    
        \multicolumn{1}{r}{}   & $\sigma=0.5$ & $\sigma=1.0$ & $\sigma=1.5$ \\ 
        $\mathcal{C}$ & \DPparam{10.87}{$10^{-4}$} & \DPparam{2.216}{$10^{-4}$} & \DPparam{0.935}{$10^{-4}$}\\ 
    \midrule
        50 & \colorbox{gray!40}{0.235/0.879/0.816} & \colorbox{gray!10}{0.190/0.876/0.846} & \colorbox{gray!10}{0.190/0.883/0.733} \\
        100 & \colorbox{gray!10}{0.180/0.887/0.775} & \colorbox{gray!10}{0.200/0.874/0.784} & \colorbox{gray!10}{0.205/0.864/0.818} \\
        200 & \colorbox{gray!10}{0.215/0.858/0.822} & \colorbox{gray!10}{0.200/0.876/0.802} & \colorbox{gray!10}{0.180/0.879/0.835} \\
    \bottomrule
        \\[-0.3em]
        \multicolumn{2}{r}{(b) \spectral} & \multicolumn{2}{l}{w. \nodp : 0.245 / 0.785 / 0.766 }\\
        \\[-0.3em]
     \end{tabular}
     }
      
      \label{tab:compare_dp_main}
    \end{minipage}
    \hfill

  \end{figure*}

In this section, we investigate differential privacy on gradients as a defense method against fingerprinting attacks. Specifically, we discuss client-side differential privacy in federated learning. 

\subsection{Differential Privacy}

Differential Privacy (DP) is a framework for capturing privacy-preserving properties of a mechanism~\cite{dwork2011firm}. 
\begin{definition} [Differential Privacy]

A mechanism $\mathcal{M}: \mathcal{D} \rightarrow \mathcal{R}$ with range $\mathcal R$ and domain $\mathcal D$ satisfies $(\varepsilon, \delta)$ differentially privacy, if for any two neighboring datasets $d, d' \in \mathcal{D}$ and for any subsets $\mathcal S \subseteq \mathcal{D}$ it holds that
\begin{equation}
    \mathbb{P}[(\mathcal{M}(d) \in \mathcal S)] \leq e^\varepsilon \cdot \mathbb{P} [(\mathcal{M}(d') \in \mathcal S)] + \delta
\end{equation}
\end{definition}
Informally, the definition captures that changes in a dataset (\eg presence or absence of an individual) do not significantly change the output of a DP mechanism, and the changes are bounded by parameters $\epsilon$ and $\delta$~\cite{dwork2006our}.\footnote{Parameter $\delta$ is preferably smaller than $1/|\mathcal D|$, where $|\mathcal D|$ indicates the size of the dataset. In the experiments we set $\delta$ to $10^{-4}$.}

We adopt Differential Privacy for Stochastic Gradient Decent (DP-SGD)~\cite{abadi2016deep} at every client. That is each client performs DP-SGD locally on their dataset. The algorithm includes two main steps: \\
1) \textbf{Clipping the gradients:}
\begin{equation}
    \Bar{\theta}(\vs_i) \leftarrow \theta(\vs_i) / \max\Big(1, \frac{\|\theta(\vs_i)\|}{\mathcal{C}}\Big)
\end{equation}
2) \textbf{Adding noise to gradients:}
\begin{equation}
    \Bar{\theta} \leftarrow \frac{1}{L} \sum_{i} \Bar{\theta}(\vs_i)+ \mathcal{N}(0, \sigma^2 \mathcal{C}^2 \mI)
\end{equation}
where clipping bound $\mathcal C$ and noise multiplier $\sigma$ can be chosen to balance the extent of privacy conferred versus the efficacy of the trained model. The overall parameters of DP-SGD are calculated according to~\cite{abadi2016deep}. Note that the DP-SGD step is performed at each client after performing iterative local updates. Clipped and noisy gradients are then communicated to the server or the anonymizer. Note that since the server is not trusted, differential privacy is performed at the client.

\subsection{Defense Experiments}

We study the effects of DP on gradients by varying the noise multiplier $\sigma$ and clipping bound $C$. The target $\delta$ is set to $10^{-4}$. The main results on \newsshort are reported in Table~\ref{tab:compare_dp_main} and full results on both datasets are provided in Appendix~\ref{appendix:dp_setting}. We vary the parameters of DP, up to the strongest setting $\varepsilon = 0.935$ which corresponds to noise multiplier $\sigma=1.5$. As expected, the clustering performance of both \spectral and \greedy is negatively correlated with the privacy budget, indicating the strong correlation between the extent of differential privacy and the capability of preventing fingerprinting attacks on federated learning.

\begin{figure}[h]
\includegraphics[width=1.0\linewidth]{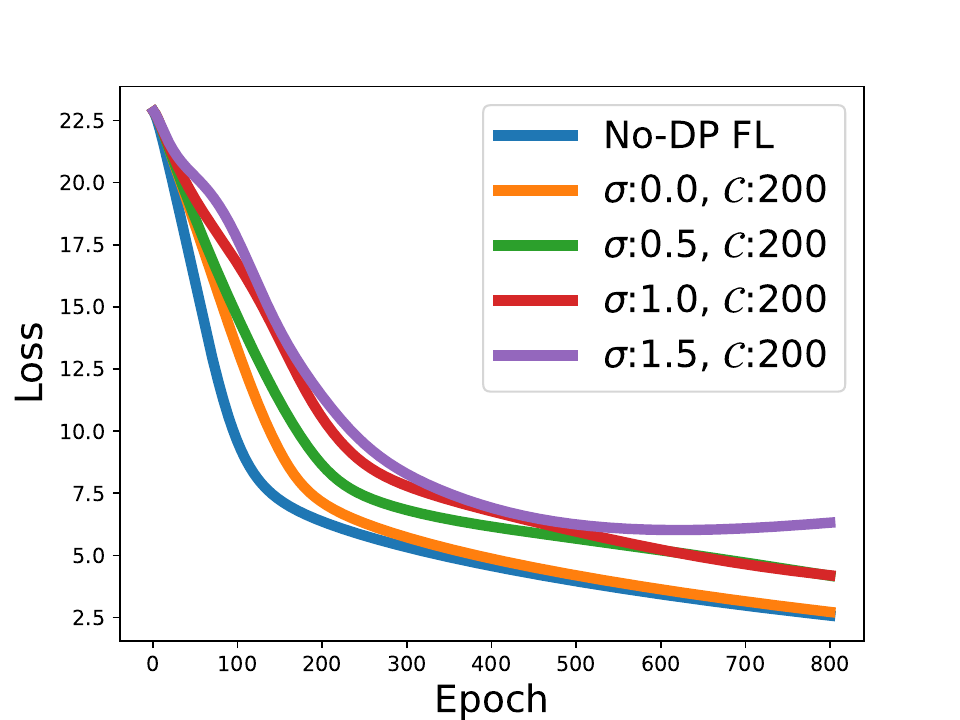}
    \captionof{figure}{The comparison of various defense methods including clipping-only ($\sigma=0$), DP-SGD (with varying noise parameters $\sigma$) and the baseline, \nodp FL, based on federated language model training on \newsfull. The loss on a holdout validation set is reported. }
    \label{fig:loss_dp_news}
\end{figure}

We further investigate the consequence of using DP-SGD in federated learning of language models. We study the vanilla FL (\nodp FL) and various DP-SGD settings on FL by comparing the model losses (log perplexity) on validation sets, as illustrated in Figure~\ref{fig:loss_dp_news}. We observe that some of these models have low utility, particularly the model with the strongest DP guarantee with $\sigma=1.5$, which begins to diverge after 500 epochs. In the setting where only clipping is used and no noise is added, $\sigma=0.0$, the model achieves a similar loss to the \nodp model, albeit with no DP guarantees. Increasing $\sigma$ to $0.5$ or $1.0$ results in about 3 times slower convergence iterations and 4 times slower running speed for each epoch, but still attains reasonable generalization performance.

\section{Conclusion}
In this paper, we evaluate privacy guarantees provided by a shuffle module if deployed in Federated Learning setting. To this end, we design a new fingerprinting attack that can link shuffled data updates across training epochs back to the same user. Our experimental results show the feasibility of our attack when training language models on shuffled gradients. DP on gradients is examined to show its effectiveness in defending against the new fingerprinting attack.

\section*{Limitations}

Our work proposes an attack technique on the Shuffle module in Federated Learning, which could be used for malicious purposes. However, the main purpose of our work is to expose the threat to the research community. We have also discussed DP as a possible defense method against the new attack. 
Though differential privacy can alleviate the fingerprinting attack, it decreases the performance of federated learning. Finding methods with a better privacy-utility trade-off is still an open question. 
Methods based on Multi-Party Computation (MPC)~\cite{bonawitz2017practical} can also be seen as defense mechanisms against fingerprinting attacks since clients send their updates in an encrypted form. However, these methods require other considerations in practice such as computational costs and participants strictly following the protocol.

Subsampling is also a widely used technology in federated learning. It may diminish the performance of \greedy as the one-to-one alignment constraint does not hold in this case. However, the clustering-based attacks, \ie \kmean and \spectral, will have less of an effect, as they do not impose the requirement of an equal number of participants for each round. 

The fingerprinting attack requires the attacker to record all gradient updates of all clients, which can lead to significant storage costs. The cost can be reduced with a few efforts: i) using Step-wise Greedy Match, which only requires storing gradients from the recent two epochs, and ii) using a single transformer layer instead of all layers or even a part of a layer, \ie FC and Proj layers, as demonstrated in Tables~\ref{tab:main_feature_combine} and ~\ref{tab:compare_layer}.

\section*{Acknowledgement}
This research was supported by The University of Melbourne’s Research Computing Services and the Petascale Campus Initiative. This work was supported in part by an Oracle Research Grant.

\bibliography{refs}

\begin{thebibliography}{10}

\bibitem{abadi2016deep}
Martin Abadi, Andy Chu, Ian Goodfellow, H~Brendan McMahan, Ilya Mironov, Kunal
  Talwar, and Li~Zhang, `Deep learning with differential privacy', in {\em
  Proceedings of the 2016 ACM SIGSAC conference on computer and communications
  security}, pp. 308--318, (2016).

\bibitem{bagdasaryan2020backdoor}
Eugene Bagdasaryan, Andreas Veit, Yiqing Hua, Deborah Estrin, and Vitaly
  Shmatikov, `How to backdoor federated learning', in {\em International
  Conference on Artificial Intelligence and Statistics}, pp. 2938--2948. PMLR,
  (2020).

\bibitem{bittau2017prochlo}
Andrea Bittau, {\'U}lfar Erlingsson, Petros Maniatis, Ilya Mironov, Ananth
  Raghunathan, David Lie, Mitch Rudominer, Ushasree Kode, Julien Tinnes, and
  Bernhard Seefeld, `Prochlo: Strong privacy for analytics in the crowd', in
  {\em Proceedings of the 26th symposium on operating systems principles}, pp.
  441--459, (2017).

\bibitem{blanchard2017machine}
Peva Blanchard, El~Mahdi El~Mhamdi, Rachid Guerraoui, and Julien Stainer,
  `Machine learning with adversaries: Byzantine tolerant gradient descent',
  {\em Advances in Neural Information Processing Systems}, {\bf 30}, (2017).

\bibitem{bonawitz2017practical}
Keith Bonawitz, Vladimir Ivanov, Ben Kreuter, Antonio Marcedone, H~Brendan
  McMahan, Sarvar Patel, Daniel Ramage, Aaron Segal, and Karn Seth, `Practical
  secure aggregation for privacy-preserving machine learning', in {\em
  proceedings of the 2017 ACM SIGSAC Conference on Computer and Communications
  Security}, pp. 1175--1191, (2017).

\bibitem{chen2019federated}
Mingqing Chen, Ananda~Theertha Suresh, Rajiv Mathews, Adeline Wong, Cyril
  Allauzen, Fran{\c{c}}oise Beaufays, and Michael Riley, `Federated learning of
  n-gram language models', in {\em Proceedings of the 23rd Conference on
  Computational Natural Language Learning (CoNLL)}, pp. 121--130, (2019).

\bibitem{criado2022non}
Marcos~F Criado, Fernando~E Casado, Roberto Iglesias, Carlos~V Regueiro, and
  Sen{\'e}n Barro, `Non-iid data and continual learning processes in federated
  learning: A long road ahead', {\em Information Fusion}, {\bf 88},  263--280,
  (2022).

\bibitem{dayan2021federated}
Ittai Dayan, Holger~R Roth, Aoxiao Zhong, Ahmed Harouni, Amilcare Gentili,
  Anas~Z Abidin, Andrew Liu, Anthony~Beardsworth Costa, Bradford~J Wood,
  Chien-Sung Tsai, et~al., `Federated learning for predicting clinical outcomes
  in patients with covid-19', {\em Nature medicine}, {\bf 27}(10),  1735--1743,
  (2021).

\bibitem{dwork2011firm}
Cynthia Dwork, `A firm foundation for private data analysis', {\em
  Communications of the ACM}, {\bf 54}(1),  86--95, (2011).

\bibitem{dwork2006our}
Cynthia Dwork, Krishnaram Kenthapadi, Frank McSherry, Ilya Mironov, and Moni
  Naor, `Our data, ourselves: Privacy via distributed noise generation', in
  {\em Annual international conference on the theory and applications of
  cryptographic techniques}, pp. 486--503. Springer, (2006).

\bibitem{erlingsson2019amplification}
{\'U}lfar Erlingsson, Vitaly Feldman, Ilya Mironov, Ananth Raghunathan, Kunal
  Talwar, and Abhradeep Thakurta, `Amplification by shuffling: From local to
  central differential privacy via anonymity', in {\em Proceedings of the
  Thirtieth Annual ACM-SIAM Symposium on Discrete Algorithms}, pp. 2468--2479.
  SIAM, (2019).

\bibitem{geiping2020inverting}
Jonas Geiping, Hartmut Bauermeister, Hannah Dr{\"o}ge, and Michael Moeller,
  `Inverting gradients-how easy is it to break privacy in federated learning?',
  {\em Advances in Neural Information Processing Systems}, {\bf 33},
  16937--16947, (2020).

\bibitem{girgis2021shuffled}
Antonious Girgis, Deepesh Data, Suhas Diggavi, Peter Kairouz, and
  Ananda~Theertha Suresh, `Shuffled model of differential privacy in federated
  learning', in {\em International Conference on Artificial Intelligence and
  Statistics}, pp. 2521--2529. PMLR, (2021).

\bibitem{goodfellow2016deep}
Ian Goodfellow, Yoshua Bengio, and Aaron Courville, {\em Deep learning}, MIT
  press, 2016.

\bibitem{gupta2022recovering}
Samyak Gupta, Yangsibo Huang, Zexuan Zhong, Tianyu Gao, Kai Li, and Danqi Chen,
  `Recovering private text in federated learning of language models', {\em
  Advances in Neural Information Processing Systems}, {\bf 35},  8130--8143,
  (2022).

\bibitem{kairouz2021advances}
Peter Kairouz, H~Brendan McMahan, Brendan Avent, Aur{\'e}lien Bellet, Mehdi
  Bennis, Arjun~Nitin Bhagoji, Kallista Bonawitz, Zachary Charles, Graham
  Cormode, Rachel Cummings, et~al., `Advances and open problems in federated
  learning', {\em Foundations and Trends{\textregistered} in Machine Learning},
  {\bf 14}(1--2),  1--210, (2021).

\bibitem{konecnyfederated}
Jakub Konecn{\`y}, H~Brendan McMahan, and Daniel Ramage, `Federated
  optimization: Distributed optimization beyond the datacenter'.

\bibitem{kuhn1955hungarian}
Harold~W Kuhn, `The hungarian method for the assignment problem', {\em Naval
  research logistics quarterly}, {\bf 2}(1-2),  83--97, (1955).

\bibitem{lamport1982byzantine}
Leslie Lamport, Robert Shostak, and Marshall Pease, `The byzantine generals
  problem', {\em ACM Transactions on Programming Languages and Systems}, {\bf
  4}(3),  382--401, (1982).

\bibitem{lang1995newsweeder}
Ken Lang, `Newsweeder: Learning to filter netnews', in {\em Machine Learning
  Proceedings 1995},  331--339, Elsevier, (1995).

\bibitem{liu2021flame}
Ruixuan Liu, Yang Cao, Hong Chen, Ruoyang Guo, and Masatoshi Yoshikawa, `Flame:
  Differentially private federated learning in the shuffle model', in {\em
  Proceedings of the AAAI Conference on Artificial Intelligence}, volume~35,
  pp. 8688--8696, (2021).

\bibitem{lloyd1982least}
Stuart Lloyd, `Least squares quantization in pcm', {\em IEEE transactions on
  information theory}, {\bf 28}(2),  129--137, (1982).

\bibitem{lyu2020differentially}
Lingjuan Lyu, Xuanli He, and Yitong Li, `Differentially private representation
  for nlp: Formal guarantee and an empirical study on privacy and fairness', in
  {\em Findings of the Association for Computational Linguistics: EMNLP 2020},
  pp. 2355--2365, (2020).

\bibitem{manning2008introduction}
Christopher~D Manning, {\em Introduction to information retrieval}, Syngress
  Publishing,, 2008.

\bibitem{mcmahan2017communication}
Brendan McMahan, Eider Moore, Daniel Ramage, Seth Hampson, and Blaise~Aguera
  y~Arcas, `Communication-efficient learning of deep networks from
  decentralized data', in {\em Artificial intelligence and statistics}, pp.
  1273--1282. PMLR, (2017).

\bibitem{mcmahan2016federated}
H~Brendan McMahan, Eider Moore, Daniel Ramage, and Blaise~Ag{\"u}era y~Arcas,
  `Federated learning of deep networks using model averaging', {\em arXiv
  preprint arXiv:1602.05629}, {\bf 2}, (2016).

\bibitem{melis2019exploiting}
Luca Melis, Congzheng Song, Emiliano De~Cristofaro, and Vitaly Shmatikov,
  `Exploiting unintended feature leakage in collaborative learning', in {\em
  2019 IEEE symposium on security and privacy (SP)}, pp. 691--706. IEEE,
  (2019).

\bibitem{ng2001spectral}
Andrew Ng, Michael Jordan, and Yair Weiss, `On spectral clustering: Analysis
  and an algorithm', {\em Advances in neural information processing systems},
  {\bf 14}, (2001).

\bibitem{passban-etal-2022-training}
Peyman Passban, Tanya Roosta, Rahul Gupta, Ankit Chadha, and Clement Chung,
  `Training mixed-domain translation models via federated learning', in {\em
  Proceedings of the 2022 Conference of the North American Chapter of the
  Association for Computational Linguistics: Human Language Technologies}, pp.
  2576--2586, Seattle, United States, (July 2022). Association for
  Computational Linguistics.

\bibitem{pfitzner2009characterization}
Darius Pfitzner, Richard Leibbrandt, and David Powers, `Characterization and
  evaluation of similarity measures for pairs of clusterings', {\em Knowledge
  and Information Systems}, {\bf 19}(3),  361--394, (2009).

\bibitem{radford2019language}
Alec Radford, Jeff Wu, Rewon Child, David Luan, Dario Amodei, and Ilya
  Sutskever, `Language models are unsupervised multitask learners', (2019).

\bibitem{rand1971objective}
William~M Rand, `Objective criteria for the evaluation of clustering methods',
  {\em Journal of the American Statistical association}, {\bf 66}(336),
  846--850, (1971).

\bibitem{rashkin-etal-2019-towards}
Hannah Rashkin, Eric~Michael Smith, Margaret Li, and Y-Lan Boureau, `Towards
  empathetic open-domain conversation models: A new benchmark and dataset', in
  {\em Proceedings of the 57th Annual Meeting of the Association for
  Computational Linguistics}, pp. 5370--5381, Florence, Italy, (July 2019).
  Association for Computational Linguistics.

\bibitem{sattler2020clustered}
Felix Sattler, Klaus-Robert M{\"u}ller, and Wojciech Samek, `Clustered
  federated learning: Model-agnostic distributed multitask optimization under
  privacy constraints', {\em IEEE transactions on neural networks and learning
  systems}, {\bf 32}(8),  3710--3722, (2020).

\bibitem{shokri2017membership}
Reza Shokri, Marco Stronati, Congzheng Song, and Vitaly Shmatikov, `Membership
  inference attacks against machine learning models', in {\em 2017 IEEE
  symposium on security and privacy (SP)}, pp. 3--18. IEEE, (2017).

\bibitem{vaswani2017attention}
Ashish Vaswani, Noam Shazeer, Niki Parmar, Jakob Uszkoreit, Llion Jones,
  Aidan~N Gomez, {\L}ukasz Kaiser, and Illia Polosukhin, `Attention is all you
  need', {\em Advances in neural information processing systems}, {\bf 30},
  (2017).

\bibitem{weller-etal-2022-pretrained}
Orion Weller, Marc Marone, Vladimir Braverman, Dawn Lawrie, and Benjamin
  Van~Durme, `Pretrained models for multilingual federated learning', in {\em
  Proceedings of the 2022 Conference of the North American Chapter of the
  Association for Computational Linguistics: Human Language Technologies}, pp.
  1413--1421, Seattle, United States, (July 2022). Association for
  Computational Linguistics.

\bibitem{xie2021crfl}
Chulin Xie, Minghao Chen, Pin-Yu Chen, and Bo~Li, `Crfl: Certifiably robust
  federated learning against backdoor attacks', in {\em International
  Conference on Machine Learning}, pp. 11372--11382. PMLR, (2021).

\bibitem{zhang2018survey}
Qingchen Zhang, Laurence~T Yang, Zhikui Chen, and Peng Li, `A survey on deep
  learning for big data', {\em Information Fusion}, {\bf 42},  146--157,
  (2018).

\end{thebibliography}
\bibliographystyle{acl_natbib}

\newpage
\appendix
\onecolumn

\section{Pre-trained Language Model Settings}
\label{appendix:pre_train}
We pre-train a local language model using WikiText101 for 20 epochs. A 4$\times$V100 server is used for pre-training the language model for 8 hr 23 minutes, achieving a final evaluation loss of 4.03. The detailed model and training settings are provided in Table~\ref{tab:pre_train_param} and~\ref{tab:model_param}.

\begin{table}

\parbox{.65\linewidth}{
\centering
\caption{Hyper-parameters for pre-training.}
\begin{tabular}{l c}
\toprule
Optimizer & AdamW \\
\midrule
Learning Rate & 5e-05 \\
Batch Size & 24 \\
Adam Beta1 & 0.9 \\
Adam Beta2 & 0.999 \\
Adam Epsilon & 1e-08 \\

\bottomrule
\end{tabular}
\label{tab:pre_train_param}
}
\hspace{-40mm}
\parbox{.65\linewidth}{
\centering
\caption{Hyper-parameters for language model.}
\begin{tabular}{l c }
\toprule
Model Type & GPT-2 \\
\midrule
Embedding Dimension & 192 \\
Number of Heads & 12 \\
Number of Layer & 4 \\
Attention Dropout Rate & 0.1 \\
Embedding Dropout Rate & 0.1 \\
\bottomrule
\end{tabular}
\label{tab:model_param}
}
\end{table}

\section{Ablation Study on Federated Learning Settings}
\label{appendix:fl_setting}
We compare the fingerprinting attack on federated learning with various learning rates ($\gamma$) for server training. All client learning rates are set to 0.1 in all our experiments. We involve 20 clients and train 10 epochs for both datasets. Although the gradients in \dialshort are noisier than \newsshort, the overall fingerprinting attack performance is consistently effective. A lower learning rate indicates a higher risk to the fingerprinting attack on both \dialshort and \newsshort, which means the risk of FL could increase towards the end of training when the learning rate is decayed to a very small value at this stage. The selected settings in our paper are highlighted with double daggers ($\ddagger$).
\begin{table*}[h]
\caption{The comparison of fingerprint attacks, \kmean, \spectral, and \greedy, on Federated Learning with various learning rate ($\gamma$). Purity (\PUR), Rand Index (\RI) and Mutual Information (\MI) are reported. * indicates \kmean does not converge in the experiment.}

\begin{center}
\begin{tabular}{ l ccc }

\toprule
 & \kmean & \spectral & \greedy \\
\midrule
$\gamma$ & \PUR / \RI / \MI & \PUR / \RI / \MI & \PUR / \RI / \MI\\
\midrule
1e-4 & 0.105 / 0.180 / 0.233 & 0.235 / 0.871 / 0.815 & 0.530 / 0.946 / 2.138 \\
1e-5 & N/A$^{*}$ & 0.235 / 0.870 / 0.885 & 0.840 / 0.974 / 2.500 \\
1e-6$^\ddagger$ & 0.100 / 0.138 / 0.199 & 0.270 / 0.896 / 0.958 & 1.000 / 1.000 / 2.996 \\
\bottomrule
\\[-0.3em]
\multicolumn{4}{c}{(a) \newsfull} \\
\\
\toprule
  & \kmean & \spectral & \greedy \\
\midrule
$\gamma$ & \PUR / \RI / \MI & \PUR / \RI / \MI & \PUR / \RI / \MI\\
\midrule
1e-5 & 0.190 / 0.650 / 0.568 & 0.290 / 0.879 / 1.120 & 0.330 / 0.921 / 1.299 \\
1e-6 & 0.185 / 0.606 / 0.514 & 0.260 / 0.906 / 1.083 & 0.295 / 0.919 / 1.202 \\
1e-7$^\ddagger$ & 0.305 / 0.776 / 0.976 & 0.450 / 0.911 / 1.592 & 0.500 / 0.939 / 1.761 \\
\bottomrule
\\[-0.3em]
\multicolumn{4}{c}{(b) \dialfull} \\
\end{tabular}
\end{center}

\label{tab:compare_lr}
\end{table*}

\newpage
\section{Ablation Study on Differential Privacy Settings}
\label{appendix:dp_setting}
We compare the effect of varying settings of DP-SGD on fingerprinting attack performance. We choose clipping value $\mathcal{C}\in\{50, 100, 200\}$ and noise multiplier $\sigma\in\{0, 0.5, 1.0, 1.5\}$.
\begin{table*}[h]
\begin{center}
\begin{tabular}{ c c c ccc }

\toprule
& & & \kmean & \spectral & \greedy \\
\midrule
Clip $\mathcal{C}$ & Noise $\sigma$ & \DPparam{$\epsilon$}{$\delta$} & \PUR / \RI / \MI & \PUR / \RI / \MI & \PUR / \RI / \MI\\
\midrule
\multicolumn{3}{c}{\nodp} & 0.100 / 0.138 / 0.199 & 0.245 / 0.785 / 0.766 & 1.000/ 1.000 / 2.996 \\
\midrule
50 & 0.0 & ($\infty$, $10^{-4}$) & 0.100 / 0.138 / 0.199 & 0.280 / 0.875 / 0.937 & 1.000 / 1.000 / 2.996 \\
50 & 0.5 & (10.87, $10^{-4}$) & 0.145 / 0.211 / 0.293 & 0.235 / 0.879 / 0.816 & 0.685 / 0.959 / 2.283 \\
50 & 1.0 & (2.216, $10^{-4}$) & 0.145 / 0.210 / 0.289 & 0.190 / 0.876 / 0.846 & 0.220 / 0.915 / 1.033 \\
50 & 1.5 & (0.935, $10^{-4}$) & 0.160 / 0.578 / 0.326 & 0.190 / 0.883 / 0.733 & 0.205 / 0.913 / 0.930 \\
\midrule
100 & 0.0 & ($\infty$, $10^{-4}$) & 0.100 / 0.138 / 0.199 & 0.265 / 0.864 / 0.943 & 1.000 / 1.000 / 2.996 \\
100 & 0.5 & (10.87, $10^{-4}$) & 0.145 / 0.210 / 0.289 & 0.180 / 0.887 / 0.775 & 0.215 / 0.914 / 0.990 \\
100 & 1.0 & (2.216, $10^{-4}$) & 0.145 / 0.210 / 0.288 & 0.200 / 0.874 / 0.784 & 0.225 / 0.914 / 1.007 \\
100 & 1.5 & (0.935, $10^{-4}$) & 0.170 / 0.538 / 0.359 & 0.205 / 0.864 / 0.818 & 0.210 / 0.914 / 0.988 \\
\midrule								
200 & 0.0 & ($\infty$, $10^{-4}$) & 0.100 / 0.138 / 0.199 & 0.265 / 0.877 / 0.964 & 1.000 / 1.000 / 2.996 \\
200 & 0.5 & (10.87, $10^{-4}$) & 0.145 / 0.210 / 0.289 & 0.215 / 0.858 / 0.822 & 0.210 / 0.914 / 0.971 \\
200 & 1.0 & (2.216, $10^{-4}$) & 0.175 / 0.470 / 0.383 & 0.200 / 0.876 / 0.802 & 0.245 / 0.916 / 1.087 \\
200 & 1.5 & (0.935, $10^{-4}$) & 0.165 / 0.584 / 0.373 & 0.180 / 0.879 / 0.835 & 0.245 / 0.915 / 1.054 \\
\bottomrule

\end{tabular}
\end{center}
\vspace{-3mm}
\caption{The comparison of various DP-SGD with \nodp on \newsfull.}
\label{tab:compare_dp_news_full}
\end{table*}

\begin{table*}[h]
\begin{center}
\begin{tabular}{ c c c ccc }

\toprule
& & & \kmean & \spectral & \greedy \\
\midrule
Clip $\mathcal{C}$ & Noise $\sigma$ & \DPparam{$\epsilon$}{$\delta$} & \PUR / \RI / \MI & \PUR / \RI / \MI & \PUR / \RI / \MI\\
\midrule
\multicolumn{3}{c}{\nodp} & 0.305 / 0.776 / 0.976 & 0.450 / 0.911 / 1.592 & 0.500 / 0.939 / 1.761 \\
\midrule
50 & 0.0 & ($\infty$, $10^{-4}$) & 1.000 / 1.000 / 2.996 & 1.000 / 1.000 / 2.996 & 1.000 / 1.000 / 2.996 \\
50 & 0.5 & (12.67, $10^{-4}$) & 1.000 / 1.000 / 2.996 & 1.000 / 1.000 / 2.996 & 1.000 / 1.000 / 2.996 \\
50 & 1.0 & (3.095, $10^{-4}$) & 0.325 / 0.793 / 1.144 & 0.740 / 0.940 / 2.239 & 0.910 / 0.987 / 2.812 \\
50 & 1.5 & (1.426, $10^{-4}$) & 0.175 / 0.483 / 0.425 & 0.290 / 0.896 / 1.087 & 0.295 / 0.919 / 1.238 \\
\midrule								
100 & 0.0 & ($\infty$, $10^{-4}$) & 0.950 / 0.993 / 2.913 & 0.510 / 0.860 / 1.972 & 1.000 / 1.000 / 2.996 \\
100 & 0.5 & (12.67, $10^{-4}$) & 0.645 / 0.935 / 2.067 & 0.860 / 0.973 / 2.586 & 0.980 / 0.997 / 2.946 \\
100 & 1.0 & (3.095, $10^{-4}$) & 0.150 / 0.235 / 0.315 & 0.205 / 0.895 / 0.854 & 0.240 / 0.915 / 1.054 \\
100 & 1.5 & (1.426, $10^{-4}$) & 0.170 / 0.577 / 0.357 & 0.205 / 0.896 / 0.876 & 0.225 / 0.915 / 1.047 \\
\midrule								
200 & 0.0 & ($\infty$, $10^{-4}$) & 0.715 / 0.940 / 2.457 & 0.795 / 0.961 / 2.632 & 0.950 / 0.995 / 2.926 \\
200 & 0.5 & (12.67, $10^{-4}$) & 0.190 / 0.568 / 0.499 & 0.230 / 0.898 / 0.958 & 0.250 / 0.916 / 1.057 \\
200 & 1.0 & (3.095, $10^{-4}$) & 0.190 / 0.638 / 0.561 & 0.225 / 0.889 / 0.934 & 0.255 / 0.916 / 1.097 \\
200 & 1.5 & (1.426, $10^{-4}$) & 0.210 / 0.624 / 0.628 & 0.280 / 0.905 / 1.097 & 0.295 / 0.919 / 1.193 \\
\bottomrule

\end{tabular}
\end{center}
\vspace{-3mm}
\caption{The comparison of various DP-SGD with \nodp on \dialfull.}
\label{tab:compare_dp_dial_full}
\end{table*}

\end{document}